\documentstyle[aps,prl,amsmath,amssymb,amscd,epsfig,subfigure,tabularx,preprint]{revtex}

\def\beq{\begin{eqnarray}}
\def\eeq{\end{eqnarray}}

\newcommand{\ck}[1]{$\check{\text{#1}}$}
\newcommand{\mf}[1]{{\mathfrak #1}}
\newcommand{\ul}[1]{\underline{#1}}

\tightenlines
\begin{document}
\title{Quantum creation of an Inhomogeneous universe}
\author{Sigbj\o rn Hervik\footnote{e-mail:sigbjorn.hervik@fys.uio.no} \\
Department of Physics, University of Oslo\\
P.O.Box 1048 Blindern\\
N-0316 Oslo, Norway}
\maketitle

\begin{abstract}
In this paper we study a class of  inhomogeneous cosmological models 
which is a modified version of what is usually called the 
Lema\^itre-Tolman model. We assume that we have a space with 
2-dimensional locally homogeneous spacelike surfaces. In addition we 
assume they are compact. Classically  we investigate both homogeneous and 
inhomogeneous spacetimes which this model describe. For instance one is a 
quotient of the AdS$_4$ space which resembles the BTZ black hole in 
AdS$_3$. 

Due to the complexity of the model we indicate a simpler model which can 
be quantized easily. This model still has the feature that it is in 
general inhomogeneous. How this model could describe a
spontaneous creation of a universe through a tunneling event is emphasized. 
\end{abstract}


\section{Introduction}
In the eighties several authors\cite{vil82,vil83,vil84,vil86,vv,AP82} suggested that 
the universe could have been  spontaneously created though a tunneling 
event. What the nature of such a tunneling event could be is still 
unsettled. Today's theories suggest however that shortly after the Planck 
era the universe was in an arbitrary state. A homogeneous and isotropic 
universe is believed to be a rather special configuration, so an 
arbitrary state suggests the universe was inhomogeneous. Recently the 
question of topology\footnote{See for instance \cite{ll}.} has been 
added into the discussion. What is the most probable topology of the 
universe? We know that there exist a lot more hyperbolic manifolds than 
elliptic and flat ones \cite{thur97}. Do we therefore live in a 
hyperbolic universe? 

In this paper we shall study a class of cosmological models which has 
2-dimensional  locally homogeneous compact spacelike surfaces. The 
2-dimensional compact oriented surfaces can be classified in term of an 
integer $g$, the {\em genus} of the surface. There is one elliptic 
surface $g=0$, one flat $g=1$ and a infinite of hyperbolic surfaces  
$g\geq 2$.

We write the metric  in a similar form to that of Kucha\ck{r}~
\cite{Kuchar} and Kenmoku and collaborators (KKTY)~\cite{KKTYa}\footnote{
Note a slightly different notation compared to the notation used by KKTY 
and Kucha\ck{r}}:
\begin{equation} \label{LTmetric}
ds^2=-N^2dt^2+Q^2(dr+N^r dt)^2+R^2 d\Omega_g^2
\end{equation}
$N, N^r, Q$ and $R$ are all functions of $r$ and $t$, and $d\Omega_g^2$ 
is the metric the $g$-holed torus\footnote{Or more correctly: 
$d\Omega_g^2$ is {\em a} metric on the $g$-holed torus, $T_g$. Since the 
Teichm\"uller space for $g\geq 1$ have dimension greater than 1, there 
are an uncountable number of non-isometric metrics on $T_g,~g\geq 1$.}, 
$T_g$. These metrics cover a great variety of different spatial 
topologies. In terms of the Thurston classification\cite{thurston}, their 
covering spaces are for instance: 
$E^3,~S^3,~{\mathbb H}^3, ~{\mathbb R}\times S^2$ 
and  ${\mathbb R}\times {\mathbb H}^2$. 
Thus a great variety of cosmological models 
are incorporated in the metric (\ref{LTmetric}). Our main concern will 
however be the Lema\^itre-Tolman models (or LT for short). In the LT 
models $g=0$. This case includes the first three of the Thurston 
geometric models. The case where $R'=Q'=N'=0$ are the Kantowski-Sachs 
($g=0$) metric and their hyperbolic versions $(g\geq 2)$ studied by 
Fugandes\cite{fugandes82}. Throughout this paper we will denote all the 
cases ${\mathbb R}\times T_g$ as KS. This model will link the models 
studied in the paper by Nambu and Sasaki\cite{NS} with for instance  the 
homogeneous FRW models or the Einstein deSitter models. This model can 
therefore among others describe a quantum tunneling of a universe from a 
Schwarzschild Black Hole into a deSitter universe.

We will choose the curvature of $T_g$ to be $\kappa=1,0,-1$ for 
$g=0,~g=1$ and $g\geq 2$ respectively. The Gauss-Bonnet Theorem then 
implies
\beq 
{\cal A}_g \kappa =4\pi(1-g)
\eeq
where ${\cal A}_g$ is the area of the $g-$holed torus $T_g$. If we choose 
${\cal A}_1=4\pi$ we can write 
\beq 
{\cal A}_g =4\pi n_g
\eeq
where 
\beq 
n_g=\begin{cases}
1, & g=0, 1 \\
(g-1), & g\geq 2
\end{cases}
\eeq

From the action of pure gravity\footnote{In this paper we have used the 
convention: $c=G_N=\hbar=k_B=1$.}, $S_G=\frac{1}{16\pi}\int 
d^4x(R-2\Lambda)$, we obtain for these models
\begin{align}\begin{split}
S_G=\int dt\int & drn_g\bigg{[} -N^{-1}\left(R(-\dot{Q}+(Q N^r)')
(-\dot{R}+R'N^r) 
 +\frac{1}{2}Q(-\dot{R}+R'N^r)^2\right)\\ 
&+N\left(-Q^{-1}RR''-\frac{1}{2}Q^{-1}{R'}^2+Q^{-2}RR'Q' 
+\frac{1}{2}Q(\kappa-\Lambda R^2)\right)\bigg{]}
\end{split}\end{align}
where the prime and dot denote the derivative with respect to $r$ and $t$ 
respectively. A cosmological constant is also included.
We will also consider a matter term or source term of the form:
\beq
S_D=-4\pi n_g\int dt \int dr N\tilde{\rho}
\eeq
where $\tilde{\rho}$ an independent function. We introduce this source 
function because it is one of the simplest ways to generate 
inhomogeneities. In the presence of 
classical dust this function   $\tilde{\rho}$ could be related to the 
dust density $\rho$ by the equation $\tilde{\rho}=QR^2\rho$. Thus in 
this description $\tilde{\rho}$ is not a dynamical matter field, we have 
basically solved the matter equations and reinserted the solutions into 
the Lagrangian\footnote{If we want to treat dust as a dynamical field we 
get a term in the Lagrangian that is linear in the conjugated momentum of
 the dust field. See for instance \cite{KT,salopek}.} .

The total action is now assumed to be $S=S_G+S_D$. 
The canonical momenta are:
\begin{align}
P_Q= & -n_g\frac{R}{N}(\dot{R}-R'N^r) \label{canmomQ} \\
P_R= & -n_g\frac{1}{N}[R(\dot{Q}-(QN^r)')+Q(\dot{R}-R'N^r)]\label{canmomP}
\end{align}
Through a Legendre transform we can bring the action into a canonical 
form:
\[
S=\int dt \int dr[ P_R\dot{R}+P_Q\dot{Q}-(N{\mathfrak H}
+N^r{\mathfrak H}_r)]
\]
where 
\begin{align}\begin{split}
{\mathfrak H}= & -\frac{1}{n_gR}P_QP_R+\frac{1}{2n_g}\frac{Q}{R^2}P_Q^2 \\
&+n_g\left[\frac{RR''}{Q}-\frac{RR'Q'}{Q^2} 
 +\frac{1}{2}\frac{{R'}^2}{Q}-\frac{1}{2}\kappa Q+\frac{1}{2}\Lambda Q R^2
+4\pi\tilde{\rho} \right]\label{enconstraint}
\end{split}\\
{\mathfrak H}_r= & R'P_R-QP_Q' \label{momconstraint}
\end{align}

\section{The constraints in the KSLT model}

From the canonical form of the action we can readily write down the energy
 constraint and the momentum constraint:
\[{\mathfrak H}=0, \quad {\mathfrak H}_r=0 \]

It is also useful to study a special combination of these two constraints,
\begin{equation}
-\frac{R'}{Q}{\mathfrak H}-\frac{P_Q}{RQ}{\mathfrak H}_r=n_g(M'-m')=0 
\label{masscons}
\end{equation}
where $M$ is called the mass function and is defined by
\begin{equation*} M\equiv -\frac{1}{2}\left[-\frac{P_Q^2}{Rn_g^2}
+\frac{R{R'}^2}{Q^2}-\kappa R+\frac{\Lambda}{3}R^3\right] 
\end{equation*}
and $m$ is defined by
\[ m\equiv \int^r dr 4\pi \tilde{\rho}\frac{R'}{Q} \]
Inserting the relation $\tilde{\rho}=\rho R^2Q$ we see that in the case 
$\kappa=1$ we can interpret the function $m$ as the mass of the dust 
inside a spherical shell of radius $R$. 

If we calculate the equal-time Poisson algebra we obtain:
\beq
\{ \mf{H}_r (r),\mf{H}_r(r^*)\}&=&\mf{H}_r(r)\delta'(r-r^*)-\mf{H}_r(r^*)
\delta'(r^*-r)\\
\{ \mf{H} (r),\mf{H}_r(r^*)\}&=&\mf{H}'(r)\delta(r-r^*)+\mf{H}(r)
\delta'(r-r^*)\\
\{ \mf{H}_r (r),\mf{H}_r(r^*)\}&=&\mf{H}_r(r)\delta'(r-r^*)-\mf{H}_r(r^*)
\delta'(r^*-r)\\
\{ \mf{H} (r),\mf{H}(r^*)\}&=&Q(r)^{-2}\mf{H}_r(r)\delta'(r-r^*)
-Q(r^*)^{-2}\mf{H}_r(r^*)\delta'(r^*-r)\\
\{ M(r),\mf{H}_r(r^*)\}&=&M'(r)\delta(r-r^*)\\
\{ M(r),\mf{H}(r^*)\}&=&-Q(r)^{-3}R'(r)\mf{H}_r(r)\delta(r-r^*)\\
\{M(r),M(r^*)\}&=&0
\eeq
Thus according to the above equations the mass function is a constant of 
motion in a weak sense.
\section{Classical solutions}
For simplicity, let us work in the gauge $N^r=0$. But for the time being 
$N$ is still arbitrary. 
The constraint ${\mathfrak H}_r$ can easily be solved in quadrature. 
Using the expressions for the canonical momenta, eq. \ref{canmomQ} and 
\ref{canmomP},  combined with the constraint equation \ref{momconstraint},
 and simplifying yields the differential equation:
\[ R'\dot{Q}-Q\dot{R}'+\dot{R}Q\frac{N'}{N}=0 \]
which has the solution $R'=FQ$. The function $F$ satisfies
\[ R'\dot{F}-F{\dot{R}}\frac{N'}{N}=0\]
For the present time we have to be a bit careful when we solve this 
equation. If $R'=0$ which happens in the KS case, the momentum constraint 
yields $N'=0$. 
For $R'\neq 0$ we can solve the equation for $F$:
\[ F(r,t)=f(r)\exp\left[\int^t\left(\frac{\dot{R}}{R'}\frac{N'}{N}\right) 
dt \right] \]
and $f(r)$ is an arbitrary function of $r$. We can add these results 
together and keep the above expression for $F(r,t)$ but with the 
additional restriction $f(r)=0$ whenever $R'=0$. We will ignore the 
possibility that the term $\frac{N'}{R'}$ in the exponential function  
causing trouble\footnote{We get a ``$\frac{0}{0}$''-expression for  
$\frac{N'}{R'}$.}. Thus if there exists a $t^*$ and a $r^*$ such that 
$R'(t^*,r^*)=0$, then $F(t,r^*)=0$ for all $t$. 

The mass function is always well defined and turns into:
\beq\label{meq}
M=\frac{R\dot{R}^2}{2N^2}+\frac{1}{2}(\kappa -F^2)R-\frac{\Lambda}{6}R^3=m
\eeq
where $m(r)=\int^r dr 4\pi \tilde{\rho}F$. If $\dot{F}=0$ the mass 
equation (\ref{meq}) is separable and can be solved in quadrature.

\subsection{Time independent solutions}
Let us assume  $\dot{R}=0$. This implies $\dot{F}=0$. Solving the mass 
equation with respect to $F$ yields:
\[ F^2=\kappa-\frac{2m(r)}{R}-\frac{\Lambda}{3}R^2 \]
In order to find the lapse $N$ we have to use the equation for 
$\dot{P}_R$:
\beq
-\frac{1}{N}\dot{P}_Q &=& \frac{1}{N}\{P_Q,\int drN\mf{H}\} \\
&=& \frac{1}{2n_g}\frac{P^2_Q}{R^2}+ n_g\left(\frac{1}{2}
\frac{{R'}^2}{Q^2}-\frac{1}{2}(\kappa-\Lambda R^2)+\frac{N'RR'}{NQ^2}\right)
\eeq
Inserting $F^2$ into the equation for $P_Q$ yields the following equation 
for the lapse $N$:
\begin{equation}\label{lapseeq}
\frac{N'}{N}=\frac{1}{2}\left(-\frac{1}{R}
+\frac{\kappa-\Lambda R^2}{\kappa R-2m-\frac{\Lambda}{3}R^3}\right)R'
\end{equation}
As we see, we might as well (at least locally) consider $R$ as our new 
radial variable, i.e. $R=r$. If we do so, the solution to the above 
equation can be written as:
\[ N^2=\eta^2(t)\frac{1}{r}e^{I} \]
where
\[ I=\int^rdr\left(\frac{\kappa-\Lambda r^2}{\kappa r-2m(r)
-\frac{\Lambda}{3}r^3}\right)\]
and $\eta(t)$ is an arbitrary function of $t$. Thus 
\beq \label{timeinmetric}
ds^2=-\eta^2(t)\frac{1}{r}e^{I}dt^2+\frac{dr^2}{\kappa-\frac{2m(r)}{r}
-\frac{\Lambda}{3}r^2 }+r^2 d\Omega_g^2 
\eeq
wherever $\kappa-\frac{2m(r)}{r}-\frac{\Lambda}{3}r^2>0$

If for instance we have constant dust density, i.e.\footnote{Note that in 
general $\hat{\rho}\neq\rho$. We  define $\hat{\rho}$ so that it is not a
 funtion of $t$, while $\rho$ in general could be a function of $t$.}
$m(r)=\mu+\frac{4}{3}\pi\hat{\rho} r^3$ the integral may be evaluated in 
terms 
of elementary functions~\cite{hovedoppg}.
\subsubsection*{Spatial topologies in AdS$_4$ space} 
\begin{figure}
\centering
\epsfig{figure=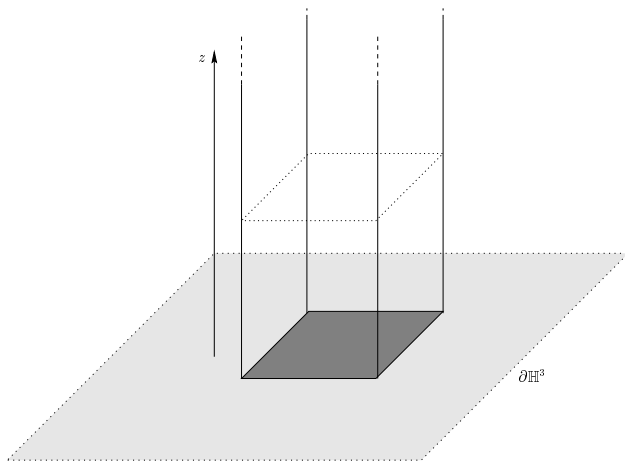, width=8.5cm} 
\caption{The space $ds^2=\frac{1}{z^2}(dz^2+d\Omega^2_1)$}\label{flat}
\end{figure}

\begin{figure}
\centering
{\epsfig{figure=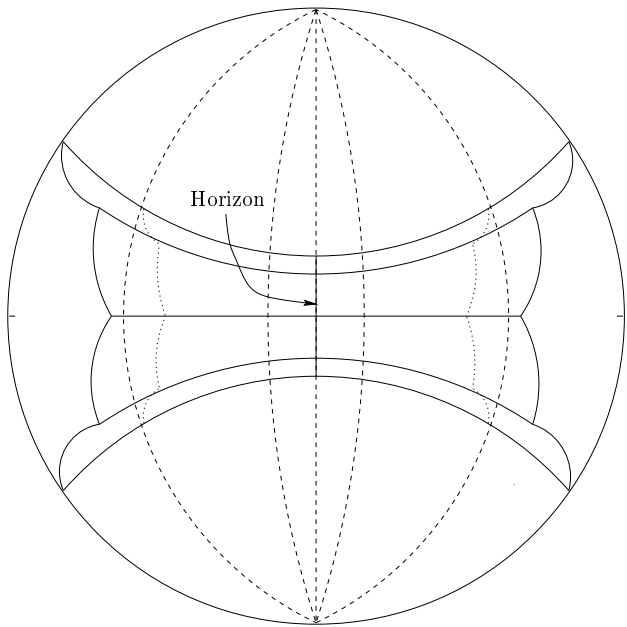, width=8.5cm} 
\caption{The AdS$_4$ apple. This figure is not drawn to scale. } 
\label{apple}}
\end{figure}

For all the deSitter spaces ($m=0$) the integral $I$ is simply:
\beq 
I=\ln\left(\kappa r-\frac{\Lambda}{3}r^3\right)
\eeq
for physical spacetimes. We notice that $\kappa\leq 0$ is not compatible 
with a $\Lambda >0$. But for $\Lambda <0$ we may have any $\kappa$. All of
 these spaces will have spatial sections which have constant negative 
curvature, i.e. they belong to the Thurston classification 
${\mathbb H}^3$.
 The cases $\kappa =0, -1$ are not compatible with compact spatial 
sections, but $\kappa=1$ may have compact spatial sections if we 
demand that the metric \ref{timeinmetric} is defined locally in $r$. If 
we demand that the metric should be globally defined then we obtain the 
space ${\mathbb H}^3$ itself. Unfortuneatly a classification of the 
possible compact quotients of ${\mathbb H}^3$ is still far from being 
established.
Let us for simplicity's sake set $\Lambda =-3$.
\begin{enumerate}
\item[$\bullet$]\ul{$\kappa=0$:} This is the non-compact multiply 
connected space investigated by Sokoloff and Starobinskii \cite{SS}. This 
space is best illustrated in the Poincar\'e half space model\footnote{See 
figure \ref{flat}.}. If we do the coordinate transformation 
$r=-\frac{1}{z}$ the metric turns into 
\[ ds^2=\frac{1}{z^2}(-dt^2+dx^2+dy^2+dz^2) \]
which is manifest on Poincar\'e half space form\footnote{Doing a Wick 
rotation of the time component, $\tau=it$, we obtain the ``Euclidean'' 
section of AdS$_4$ space. Here we see that the Euclidean section is the 
4-dimensional hyperbolic space ${\mathbb H}^4$.}. We identify points 
along two pairs of planes parallel to the $z$-axis. It is easily seen 
that this space does not allow compactification in the $z$ (or $r$) 
direction.
\item[$\bullet$]\ul{$\kappa=-1$:} This section of the hyperbolic space can
 be illustrated as follows. Draw two $4g$-gons on the boundary of the 
Poincare ball model. These polygons can be placed so that they are 
symmetric to each other through the origin. Connecting the polygons using 
geodetic planes we obtain a space that look like an apple core(see figure 
\ref{apple}). Also this case will not allow a compactification in the $r$ 
direction. Actually we have cheated a little bit, because the space 
contains a
 horizon with  topology as a $g$-torus, much like that of a black hole
\cite{blp}. The area of this horizon is 
\[
A_g=\frac{12\pi}{|\Lambda|}n_g
\]
Thus apparently we can relate to it the Bekenstein-Hawking entropy:
\[ S_{BH}=\frac{1}{4}\int_{T_g}dA=\frac{3\pi}{|\Lambda|}n_g\]
This metric is very similar to the BTZ black hole in AdS$_3$\cite{BTZ}. 
Including a mass term in this case is straight forward. Note also that in 
the above case the ``Black Hole'' entropy is a feature of the negatively 
curved compact spaces $T_g$.  
To obtain the AdS$_4$ apple we have actually taken two such spaces and 
glued them along their horizons to obtain a topological complete space. 
Note also that the horizon is the only $r=constant$ surface which is 
totally geodesic. To ensure regularity of the total space after we have 
performed a gluing we have to glue along totally geodesic surfaces. 
\end{enumerate}

If we had demanded that all of these cases should be simply connected, 
they would have yielded the same space ${\mathbb H}^3$. But since we have 
assumed that our locally homogeneous 2-dimensional surfaces are compact 
all of the above cases yield topologically different spaces. Again 
\cite{sig} we see that assuming {\em compact} 2-dimensional locally 
homogeneous spaces ensure that the different cases will yield different 
solutions.
\subsection{Kantowski-Sachs-like solutions}
In the Kantowski-Sachs case we have $R'=0$. This implies $F=0$. The mass 
equation will now yield:
\beq 
\mu=\frac{R\dot{R}^2}{2N^2}+\frac{1}{2}\kappa R-\frac{\Lambda}{6}R^3
\eeq
where $\mu$ is a constant. Inserting this into the energy constraint we 
obtain an equation similar to the equation \ref{lapseeq}:
\beq\label{Qeq}
\frac{\dot{Q}}{Q}=\frac{1}{2}\left(-\frac{1}{R}+\frac{\Lambda R^2+8\pi
\rho R^2-\kappa}{\frac{\Lambda}{3}R^3+2\mu-\kappa R}\right)\dot{R}
\eeq
Again we see that we can equally well consider $R$ to be $R=t$. This 
equation is very similar to the equation \ref{lapseeq}. This is not 
surprising since if we smoothly continue the Schwarzschild metric to the 
region inside the black hole horizon we obtain the KS metric inside the 
horizon\cite{NS}. Smooth continuing  the deSitter solution to the 
region outside the deSitter horizon gives the KS metric\cite{BH}. We also 
note that the $t\leftrightarrow r$ symmetry is manifestly broken with the 
inclusion of classical dust. Due to the conservation of the rest mass of 
the dust, we would obtain $\int_{\Sigma}d^3x\rho Q R^2=V_c{\mathcal M}$ 
where $V_c=\int_{\Sigma}d^3x$ is the coordinate volume if the spatial 
sections $\Sigma$ are compact.

We re-write these equations for a general dust distribution. We assume 
that the topology in the $r$-direction  is that of a circle $S^1$. If 
$0\leq r \leq \xi$ then we can write 
\beq 
\tilde{\rho}={\mathcal M}+\sum_{n=1}^{\infty}\left[a_n(t)\cos\left(
\frac{2\pi nr}{\xi}\right)+a_{-n}(t)\sin\left(\frac{2\pi nr}{\xi}\right)
\right]
\eeq
We use the same trick to write $\left(\frac{\Lambda}{3}t^2+\frac{2\mu}{t}
-\kappa\right)^{-\frac{1}{2}}Q=G$ where
\beq
G(r,t)=\sum_{n=0}^{\infty}\left[c_n(t)\cos\left(\frac{2\pi nr}{\xi}\right)
+c_{-n}(t)\sin\left(\frac{2\pi nr}{\xi}\right)\right]
\eeq

Since the cosines and the sines form a complete and orthogonal set of 
functions on the circle the equations for $c_n,~n\neq 0$ are
\beq 
\dot{c}_n(t)=\frac{4\pi a_n(t)}{t\left(\frac{\Lambda}{3}t^2+\frac{2\mu}{t}
-\kappa \right)^{\frac{3}{2}}}
\eeq
The equation for $c_0(t)$ is 
\beq
\dot{c}_0(t)=\frac{4\pi{\mathcal M }}{t\left(\frac{\Lambda}{3}t^2
+\frac{2\mu}{t}-\kappa \right)^{\frac{3}{2}}}
\eeq
can be integrated at once by elementary calculus. 

In the absence of dust $\tilde{\rho}=0$ this reduces to the ordinary 
inner/outer  solutions of the Schwarzschild-deSitter spacetime. This is 
just the homogeneous solution of the KS metric. If there are 
inhomogeneities we have to solve the equations for $c_n(t),~n\neq 0$. 
Unfortunately we then have to know the evolution of the matter content. If
 we on the other hand knew the metric, i.e. the functions $c_n(t)$ we 
could uniquely determine the matter distribution.

\section{LT solutions}
So far we have only investigated time-independent and KS-like solutions. 
We saw that they represented different sections of a more general 
spacetime. Let us now look at the LT solutions, which we can obtain if we 
choose the universal time gauge $N=1$. The coordinates now follow the 
collapsing/expanding dust. 

If we choose $N=1$, $F=F(r)$ will follow. 
We can now write the mass equation as  an ``energy'' equation:
\[
\frac{1}{2}\dot{R}^2+V(r,R)=E(r)
\]
where the ``potential'' $V$ and the ``energy'' $E$ are given by
\[ V=-\frac{m}{R}-\frac{\Lambda}{6}R^2 \]
\[ E=-\frac{1}{2}(\kappa-F^2)\]

This energy equation may be integrated and solved exactly. A summary of 
the results and a qualitative description of the physical meaning of the 
solutions is given in ~\cite{Kra:97}. The solutions may be written in 
terms of the Weierstrass' elliptic functions ~\cite{Zecca:90}. The actual 
expressions are not very informative unless the reader has massive 
knowledge of these elliptic functions. However a lot of qualitative 
information can be extracted from simple classical considerations. 

We will however solve the energy equation in the big and small $R$ limit. 
\subsubsection*{Large $R$: Neglecting the $m$ term} 
If we neglect the $m$ term in the energy equation, we get a simple and 
easily solvable equation. The equation becomes:
\[ \dot{R}^2-H^2R^2=2E \]
where $H^2=\frac{\Lambda}{3}$ (assuming $\Lambda>0$) This equation can be 
solved in quadrature, and the solutions are deSitter-like 
solutions\footnote{For a rich and qualitative description of various 
deSitter models see for instance~\cite{EG:95}}:
\begin{equation}
R(r,t)=\begin{cases}
\frac{\sqrt{2E(r)}}{H}\sinh(H(t-t_0(r))), & E>0 \\
e^{H(t-t_0(r))},& E=0 \\
\frac{\sqrt{2|E(r)|}}{H}\cosh(H(t-t_0(r)),& E<0
\end{cases}
\end{equation}
\subsubsection*{Small $R$: Neglecting the $\Lambda$ term} If the $\Lambda$
 term is neglected, we end up with a energy equation which looks like:
\[ \dot{R}^2-\frac{2m}{R}=2E \]
Introducing a parameter $\Theta$ we can also solve this equation in 
quadrature:
\begin{align}
\left\{\begin{matrix} R&=&\frac{m}{2E}(\cosh \Theta-1) \\
t-t_0(r)&=&\frac{m}{(2E)^{\frac{3}{2}}}(\sinh\Theta-\Theta)
\end{matrix}\right\} &, E>0 \label{storreennull} \\
R=\left(\frac{9}{2}m\right)^{\frac{1}{3}}(t-t_0(r))^{\frac{2}{3}} &, E=0 
\label{liknull}\\
\left\{\begin{matrix} R&=&\frac{m}{2|E|}(1-\cos \Theta) \\
t-t_0(r)&=&\frac{m}{(2|E|)^{\frac{3}{2}}}(\Theta-\sin\Theta)
\end{matrix}\right\} &, E<0 \label{mindreennull}
\end{align}

From these solutions we notice that all the small $R$ solutions goes as \\
${R\approx \left(\frac{9}{2}m\right)^{\frac{1}{3}}(t-t_0(r))^{
\frac{2}{3}}}$ as {$R\longrightarrow 0$}. In the special case 
$t_0(r)=constant$ and $F^2=1-kr^2$ the equations \ref{storreennull}, 
\ref{liknull} and \ref{mindreennull} reduces to the well known matter 
dominated homogeneous FRW solutions.
\subsubsection*{The behavior of a general solution}
\begin{figure} 
\centering
\epsfig{figure=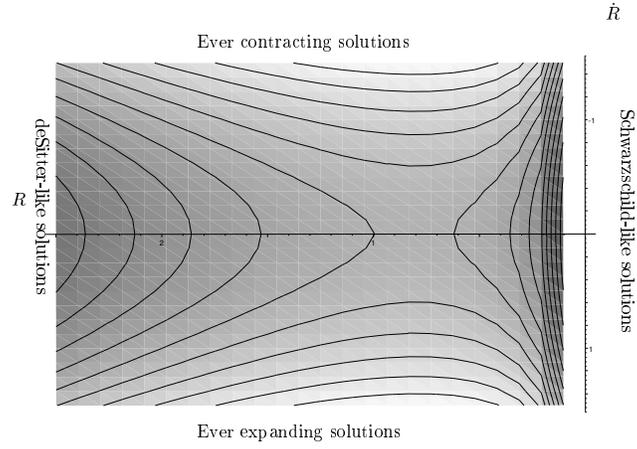, width=8.5cm}
\caption{Level curves of $E(R, \dot{R})=\frac{1}{2}\dot{R}^2-\frac{1}{5R}
-\frac{1}{5}R^2$} \label{levelcurves}
\end{figure}

The classical solutions will move on level curves of the energy function 
{$E(R, \dot{R})=\frac{1}{2}\dot{R}^2-\frac{m}{R}-\frac{\Lambda}{6}R^2$}, 
since the total energy $E$ is independent of $t$. In figure 
\ref{levelcurves} the level curves of a typical energy function are drawn.
 If $\Lambda>0$ there will exist a saddle-point of the energy function. T
his saddle point will be at $R=\left(\frac{3m}{\Lambda}\right)^{
\frac{1}{3}}, \dot{R}=0$ where the energy function will have the value 
{ $ E_s=-\frac{1}{2}(9m^2\Lambda)^{\frac{1}{3}}$}. The saddle point 
solution is  static, and is as a matter of fact the Einstein static 
universe\footnote{This is easily seen if we define the new radial variable
 to be $R$. This can be done since we have to assume that $m'(r)>0$ on 
physical grounds. In the case $m'(r)=0$, the metric become degenerate.}.
If $E<E_s$, the solutions fall into two distinct classes:
\begin{enumerate}
\item{} Schwarzschild-like solutions: These solutions expands, but they 
do not possess enough energy to escape the gravitational collapse, so they
 end as  black holes. If $m$ is constant these solutions are those of a 
Schwarzschild black hole in Lema\^itre coordinates. 
\item{} deSitter-like solutions: solutions where the universe evolves 
approximately as that of deSitter solutions with positively curved 
hypersurfaces.  
\end{enumerate}
If $E>E_s$ the (test) matter has enough energy to escape the gravitational
 collapse (expanding solutions) or enough energy to prevent the 
gravitational repulsion from the cosmological constant (contracting 
solutions). 
\subsubsection*{NOTE:} All of these solutions are also valid for the case 
$F=0$ the KS case. But to obtain the KS solutions it is not enough to 
solve just the mass equation.  

\section{The Quantum mechanical LT model}

From here on we will only consider the case where the cosmological 
constant is non-negative. 
As we saw from the classical case, if the ``energy'' is small enough the 
solutions are confined to be either Schwarzschild-like or deSitter-like. 
These two regions are separated by a classically forbidden region. From 
classical quantum mechanics we have learned that this does not have to be 
the case for a quantum system. A quantum mechanical system may tunnel 
though classical barriers. In our case this means that a small 
Schwarzscild-like universe may tunnel though the potential barrier to 
become a deSitter-like universe! This has striking consequences. If we 
imagine that a ``mother'' universe could spontaneously create 
Schwarzschild-like black holes of less than a Planck-length, there will 
be a non-zero probability that one of these will tunnel through the 
potential barrier, and end up in an ever-expanding deSitter state. These 
considerations are of course rough and based on classical quantum 
mechanics. Whether this scenario is probable in the yet unknown theory of 
Quantum Gravity remains to be proven. However we will use a simplified 
version of the WD equation and estimate the form of the wave function in 
various limits. We will from now on consider only the spherically 
symmetric case, i.e. $\kappa=1$.

\subsection{The canonical Dirac-quantization}
The Dirac quantization procedure is to let the constraint equations become
 operator equations of the wave function:
\begin{align}
\hat{\mathfrak H}\Psi =0 \\
\hat{\mathfrak H}_r\Psi =0
\end{align}
The first of these equations is the Wheeler-DeWitt equation. In addition 
we have the momentum constraint equation. These two equations are 
functional differential equations and thus in general very difficult to 
solve. KKTY managed in the case $\rho=0$ to find a couple of special cases
 where the wave function was exactly derived and interpreted 
\cite{KKTYb}. Since the mass equation is a special combination of the 
momentum and energy constraint, the mass equation itself can be 
considered as a constraint. The wave function has to obey:
\begin{equation}
\hat{M}\Psi=m\Psi
\end{equation}
The wavefunction has to be an eigenfunctional of the mass-operator 
$\hat{M}$ with eigenvalue $m(r)$, and from the classical interpretation 
we have to assume that the eigenvalue $m(r)$ has to obey $m'(r)\geq 0$. 

In the WKB approximation we can integrate the Hamilton-Jacobi equation if 
we choose a particular and simple path ~\cite{FMP:90}.  We will not use 
these results. Neither will we give any attempt to solve all three 
constraint equations simultaneously in the ``correct'' quantum-mechanical 
manner. 
To simplify the equations we will do the following:
\begin{enumerate}
\item[$\bullet$] Solve the momentum constraint classically. When we solve 
it we will assume that $R'\neq 0$, thus the KS case is excluded. Solving 
the momentum constraint will reduce the number of canonical variables, so 
we end up with only $R$ and its canonical conjugated momentum $\Pi_R$. 
\item[$\bullet$] Choose the gauge $N'=0$ so that we will not have any 
problems with the function $F$ being dependent of $t$.
\item[$\bullet$] Find the  new Hamiltonian which reproduces the classical 
equations. 
\end{enumerate}

The reduced action which reproduces the classical equations is:
\[ S=\int dt\int dr\frac{N}{F}\left[-\frac{1}{2}\frac{R\dot{R}^2}{N^2}
+\frac{1}{2}R(1-F^2)-\frac{\Lambda}{6}R^3-m(r)\right]' \]
Calculating the canonical momentum
\[ \Pi_R\equiv \frac{\delta S}{\delta \dot{R}}=\left(\frac{N}{F}\right)'
\frac{R\dot{R}}{N^2} \]
 doing a Legrendre transform to bring the action onto a canonical form, 
and removing boundary terms, we arrive at the  Hamiltonian which 
reproduces the classical equations:
\begin{equation}\label{newmodel}
H=\int dr\frac{N}{F}\left[-\frac{1}{2}\frac{F^4\Pi_R^2}{R{F'}^2}
-\frac{1}{2}R(1-F^2)+\frac{\Lambda}{6}R^3+m(r)\right]'
\end{equation}

The constraint equation will now yield the correct classical equation; 
the mass equation. Calculating the equation for $\Pi_R$, we find the 
identity, $\dot{M}=0$, which tells us that the mass function is 
independent of time.

The wave function has to satisfy the constraint equation:
\begin{equation}
\frac{\partial}{\partial r}\hat{{\mathfrak H}}\Psi=0 
\end{equation}
where
\[ \hat{{\mathfrak H}}=\frac{1}{2}\frac{F^4}{R{F'}^2}
\frac{\delta^2}{\delta R^2}-\frac{1}{2}R(1-F^2)+\frac{\Lambda}{6}R^3 
+m(r)\]

A possible constant, due to the derivative with respect to $r$ is assumed 
to be absorbed into the function $m$. 
\par
This simplified version of the LT model is not only simpler but the 
action has the great advantage that it reduces to the deSitter models 
studied by Vilenkin and Hartle-Hawking if we demand that $m=0$, $R=a(t)r$ 
and $F^2=1-kr^2$.\footnote{Inserting $m=\frac{4}{3}\pi\hat{\rho} r^3$, 
$R=a(t)r$
 and $F^2=1-kr^2$ where $\hat{\rho}$ and $k$ are constants into the 
action, the 
$r$ dependent part can be factored out, and the integral may be performed.
 The actual integral obtained is: $\int dr\frac{r^2}{\sqrt{1-kr^2}}=
\frac{\pi}{4k^{\frac{3}{2}}}$ in the case of $k>0$.} The solution most 
likely to describe the tunneling behaviour emphasized in the beginning of 
this section, is the tunneling wavefunction suggested by Vilenkin. 

\subsection{The tunneling wave function in the WKB approximation}
In the WKB approximation we assume that the wave function has the form 
$\Psi_{WKB}=\exp(\pm iS)$,
to the lowest order we get the Hamilton-Jacobi equation: 
\begin{equation}\label{energidiff}
\left(\frac{\delta S}{\delta R}\right)^2-
\frac{{F'}^2}{F^4}\left[2mR-R^2(1-F^2)+\frac{\Lambda}{3}R^4\right]=0
\end{equation}
If we assume that $S=\int \sigma(r) dr$ the resulting equation will be  
the Hamilton-Jacobi equation for a point particle with action $\sigma(r)$
($r$ is only a parameter). In the Hamilton-Jacobi equation the functional 
$S$ turn out to be the action at the classical level. Since the classical
 action can be written as an integral over $r$ the assumption 
$S=\int \sigma(r) dr$ is therefore reasonable at the lowest order WKB 
level. This separation of the action is 
also a feature of the dust, in the sense that dust does not self-interact.
 Because the matter does not self-interact the matter equations do not 
depend on neighbouring points. 
We can interpret the action $\sigma$  as the action of a point particle 
moving in a potential 
$V(R)=\frac{{F'}^2}{F^4}\left[-mR
+\frac{1}{2}(1-F^2)R^2-\frac{\Lambda}{6}R^4\right]$ with zero energy. The 
WKB wavefunction $\psi_{WKB}$ for the point particle can then be written
$\psi_{WKB}=\exp(\pm i\sigma)$. The two WKB wavefunctions can therefore 
be 
related by $\Psi_{WKB}=\exp(\int dr\ln \psi_{WKB})$. Finding first the 
wavefunction $\psi$ we can then relate its WKB approximation with 
$\Psi_{WKB}$ through  $\Psi_{WKB}=\exp(\int dr\ln \psi_{WKB})$.

We  imagine that the universe point starts off at $R=0$ and the behavior 
of the wave function of such a particle is dependent on the form of the 
potential. The zeros of the potential indicate classical turning points, 
separating classically allowed/forbidden regions. The potential $V(R)$ has
 one zero at $R=0$ and  possibly two more for positive values of $R$. 
Whether there are two, one or zero, depends on the entity:
\[ l=\frac{9m^2\Lambda}{(1-F^2)^3} \]
Iff $0<l<1$ then there will be two positive roots, $r_1$ and $r_2$. The 
region $r_1<R<r_2$ (assuming $r_1<r_2$) is a classical forbidden region, 
thus the wavefunction is exponential in some way. 
Iff $l=1$ these two roots have emerged into one, and the case $l>1$ and 
$l<0$ has no positive roots; the classical particle rolls down the slope 
to infinity. 

One case which is very instructive to look at is the case where 
$F^2=1-kr^2$ and the dust density is constant: 
$m=\frac{4}{3}\pi\hat{\rho} r^3$.
 The entity $l$ will then be a constant with respect to $r$:
\[ l=\frac{16 \pi^2 \hat{\rho}^2 \Lambda}{k^3} \]
We can interpret this quantum mechanically as if we imagine a 
spontaneously created universe ``bubble'' with high curvature, and a 
relatively small cosmological constant and dust density. It is most 
likely to vanish before it has the opportunity to enter a deSitter phase. 
Let us look at this from another point of view. Imagine that we can only 
observe ``big'' universes, universes at scales much larger than the 
Planck length, or essentially just universes that are in the expanding 
deSitter stage. If we could in some way sit outside and see the flow of 
universes in expanding deSitter stages, we would find  that most of 
the universes with low $\Lambda$ and $\hat{\rho}$ also have low 
hypersurface curvature. 

Let these thought experiments be for the moment because we do not actually
 know how to interpret the wave function of the universe or how to 
calculate it. Let us however try to estimate the behavior of the 
wavefunction is this model.

The most interesting case seems to be $0<l<1$, so let us just look at 
this case. There are essentially three different regions with different 
behavior. 
\begin{enumerate}
\item{Region I:} Small $R$. We can neglect the $R^4$-term.
\item{Region II:} Intermediate region. This is dominated by the part 
where the wave function has an exponential behavior. 
\item{Region III:} Large $R$. We can neglect the $R$-term.
\end{enumerate}

\subsubsection*{Region I}
We neglect the $R^4$-term in eq. \ref{energidiff} and complete the square 
with a new variable $\frac{F}{\sqrt{|F'|}}\lambda =(1-F^2)^{\frac{1}{4}}
R-\frac{m}{(1-F^2)^{\frac{3}{4}}}$. Then the Schr\"odinger equation for 
$\psi$  reduces to a Harmonic oscillator-like equation:
\[ \frac{\partial^2 \psi}{\partial \lambda^2}+(2p+1-\lambda^2)\psi =0\]
where $p=\frac{1}{2}\left(\frac{|F'|m^2}{F^2(1-F^2)^{\frac{3}{2}}}-1
\right)$.
The solution of this equation can be written in terms of the 
\emph{Parabolic cylinder functions} $D_p(z)$:
\[ \psi=\begin{cases}
D_p(\pm\sqrt{2}\lambda)\\
D_{-p-1}(\pm i\sqrt{2}\lambda) 
\end{cases}\]
In the case of $ p=n $ an integer, the $D_p(z)$ can be written in terms of
 the Hermite polynomials (as one would expect):
\[ D_n(z)=- 2^{-\frac{n}{2}}e^{-\frac{z^2}{4}}H_n(\frac{z}{\sqrt{2}})\]
The index $p$ can be any number, in contrast to the 1-dimensional Harmonic
 oscillator. 
\subsubsection*{Region III}
We neglect the $R$-term in eq. \ref{energidiff}. This is equivalent to 
setting
 the mass function $m$ equal to zero. We are then left with a deSitter 
space-time. The wavefunction for such space-times has been calculated 
using the WKB approximation. Most interesting to us is the tunneling wave 
function of Vilenkin. Through the rescaling: $\eta =\frac{\sqrt{|F'|}}{F}(
1-F^2)^{\frac{1}{4}}R$ we can bring the equation to the same form as that 
of Vilenkin~\cite{Vilenkin88}, with a scalar field potential $V=\frac{F^2
\Lambda}{3|F'|(1-F^2)^{\frac{3}{2}}}$. We  use this result to obtain:
\begin{equation} \label{viltun}
\psi\approx \begin{cases} \exp\left(-\frac{1}{3V}\left[1-(1-\frac{R^2
\Lambda}{3(1-F^2)})^{\frac{3}{2}}\right]\right) , & \frac{R^2\Lambda}{3(1
-F^2)}<1 \\
\exp\left(-\frac{1}{3V}\left[1+i(\frac{R^2\Lambda}{3(1-F^2)}-1)^{
\frac{3}{2}}\right]\right) , & \frac{R^2\Lambda}{3(1-F^2)}>1
\end{cases}
\end{equation}

Let us assume that $m(r)=0$ and $F^2=1-kr^2$ where $k$ is assumed to be 
positive. We will now use the tunneling solution eq. \ref{viltun} to find 
out what is most likely, the creation of a highly curved or a less curved 
space. If we calculate the ratio $\frac{\Psi^*\Psi|_{R=\tilde{R}}}{
{\Psi^*\Psi|_{R=0}}}$ where $\tilde{R}^2>>\frac{3}{\Lambda}(1-F^2)$, 
we get:
\begin{equation}
P_k\equiv\frac{\Psi^*\Psi|_{R=\tilde{R}}}{{\Psi^*\Psi|_{R=0}}}=
\exp\left[-\frac{2}{\Lambda}\int_0^1d(\sin\theta)G(\theta)\right]
\end{equation}
where $G$ is a real function independent of $k$ which satisfies 
$G(\theta)\geq 0, \forall \theta$. We thus have to conclude that to 
lowest order in the WKB approximation it is equally probable for a space 
with small spatial 
curvature to tunnel through the classical barrier than for a highly curved
 space. Comparing with our previous statement, we se that larger 
curvarture itself increases the probability amplitude. But smaller 
curvature 
increases the volume by exactly the amount so that it cancels the 
contribution from the curvature.  

\subsubsection*{Region II} Here the wave function will be of exponential 
behavior. But the actual shape is strongly dependent on the boundary 
condition. To find the wave function in this region to lowest order, 
the best is perhaps to use the function from Region I and III and cut and 
glue them together.  

\subsubsection*{Connection with other work}
Nordbury ~\cite{Norbury97,Norbury98} studied the FRW model with different 
types of perfect fluids. The tunneling picture which we have described 
here seems to agree with his results for a closed universe $(1-F^2>0)$. 
However in the case of open ($k=-1$) FRW-universes he concludes that these
 models can not have been created through the tunneling picture. From our 
considerations however, this can not be ruled out, on the contrary, open 
universes do seem to tunnel from a $R=0$ state a lot easier than closed 
ones. 

Also an article by Atkatz and Pagels~\cite{AP82} seems to indicate that 
open universes can not be created by a tunneling event. A key argument 
used in both of these articles is that the transition amplitude is 
suppressed by the infinite integral over the three-dimensional 
hypersurfaces. However, a more recent article by Coule and Martin
\cite{CM99} used the fact that there exists $k=-1$ compact spacelike 
hypersurfaces. Since they are compact they will have finite volume. Their 
conclusion is opposite of that of the previous two articles: The creation 
of a open $k=-1$ FRW universe is more likely than a closed $k=1$ universe 
through a tunneling event. Our case is however more subtle, since the 
hypersurfaces are no longer homogeneous. 

\subsection{Inhomogeneity Vs. Homogeneity: Tunneling Probabilities}
As we have investigated the WKB wavefunctions in this model, we might 
wonder whether an inhomogeneous universe is more likely to be created 
through a tunneling event than a homogeneous one. As an inhomogeneity 
measure in this model it is very useful to use the Weyl curvature tensor 
$C_{\alpha\beta\gamma\delta}$. All of the FRW models are conformally flat,
 thus for the all of the FRW model the Weyl tensor will vanish. In our 
case the opposite is also true, if $C_{\alpha\beta\gamma\delta}=0$ and 
$N'=0$ in the LT model then the spacetime is spatially homogeneous. If we 
define a mean dust density function $\bar{\rho}$ by:
\beq
m(r)=\frac{4}{3}\pi\bar{\rho}R^3
\eeq
we can write the Weyl curvature invariant in the LT model as:
\beq
C^{\alpha\beta\gamma\delta}C_{\alpha\beta\gamma\delta}
=\frac{16^2}{3}\pi^2(\bar{\rho}-\rho)^2
\eeq
Thus this space-time is homogeneous if and only if 
$\bar{\rho}\equiv \rho$.

The inhomogeneity in this model is therefore almost completely 
characterized by the mass function $m(r)$. The homogeneous mass function 
will go as $m_h(r) = K r^3$, where $K$ is a constant. Investigating only 
closed universes, we fix the constant $K$ so that 
$m_h(r_{max})= m(r_{max})$. 
The most physical interesting universes that these LT models describe 
will have a larger amount of dust consentrated in the inner regions than 
the outher regions. The most extreme inhomogeneous mass function is the 
Schwarzschild mass function, $m_{Sch}(r)=\mu=constant$. Thus it is very 
reasonable to assume that we have a mass distribution that obeys:
\beq
m_{Sch}(r)\geq m(r)\geq m_h(r)
\eeq
If we compare the WKB tunneling potential 
$ V(R)=\frac{{F'}^2}{F^4}\left[-mR
+\frac{1}{2}(1-F^2)R^2-\frac{\Lambda}{6}R^4\right]$ for these different 
mass distributions we see that we get
\beq
V_{Sch}(R)\leq V(R) \leq V_h(R)
\eeq
keeping all other parameters fixed. 
Thus the more inhomogeneous the space is the smaller the tunneling barrier
 will be. In the WKB approximation this means that the probability of the 
creation of an inhomogeneous universe is larger than for a homogeneous 
universe. More specific the spontaneous creation of a 
Schwarzschild-deSitter like universe appears to be more probable than a 
homogeneous FRW universe. We also have be aware of that the matter 
Lagrangian describes dust. Dust is an idealised form of matter that has 
no internal pressure. As the dust density becomes high, either in an 
early stage of the universe or in large local inhomogeneities this 
assumption of a pressure-free matter Langrangian becomes somewhat 
artificial and unphysical. In these situations we know that the internal 
pressure becomes very important and even quantum effects from the matter 
fields are essential in the description. In this sence we may say that 
these calculations more or less are ``valid'' only with small local 
inhomogeneities or they describe the effect from {\it Gravity itself} to 
produce inhomogeneities. This point is also emphasized by 
Conradi\cite{con1,con2}. Conradi discusses the wave function of an 
anisotropic KS space with a dust Lagrangian. He argues that since dust 
interacts only locally, in the sence that it does not self-interact, this
 local ``tunneling'' process cannot be interpreted as a tunneling effect. 
Our results should therefore perhaps be seen from a more phenomological 
point of view. The matter Lagrangian used in this paper is one of the 
most simple source field that give rise to inhomogeneities. The source 
field used is also interesting because it connects the Schwarzschild 
spacetime with the FRW spacetime in a smooth way. 

\section{Conclusion and Summary}
In a Hamiltonian description of General Relativity there has been a 
consensus among physicists and cosmologists that the issue of topology is 
essential and necessary in the description \cite{as,kodama}. This has 
also been emphasized 
in this paper. At the classical level we saw that apparently different 
cases would lead to the same spacetime if the symmetry spaces were 
infinite and simply connected. Assuming compactness and multiply 
connectedness this pathology was removed and yielded some interesting 
spacetimes.

In the Quantum mechanical case we provided a simpler model which was 
in general inhomogeneous, but removed many of the algebraic difficulties 
in the full KSLT model. It was emphasized how this model could describe a 
quantum creation of a inhomogeneous universe through quantum tunneling. 
We also showed that under some reasonable assumptions the creation of an 
inhomogeneous universe was more probable than a homogeneous FRW universe. 
In the absence of interacting matter this is very reasonable because 
there are a lot more inhomogeneous configurations than homogeneous ones. 

Even though the full theory of Quantum Gravity is far from being 
established, it seems like Quantum Cosmology {\em could} describe many of 
the features of the initial universe. In the recent decade the interest 
for topology in cosmology has expanded the application for Quantum 
Cosmology. Even in classical gravity the inclusion of non-trivial 
topologies has yielded new and surprising results. 

\section*{Acknowledgments}
I deeply appreciate the help and comments which \O yvind Gr\o n has given 
me during the work of this paper. 

I also thank Branko Steffensen for reading through the manuscript and 
making useful comments.


\begin{table} \label{summary}
\centering
\setlength{\extrarowheight}{4pt}
\begin{tabular}{| >{\large}c || >{\large}c | >{\large}c| >{\large}c |}
\hline
$\Lambda$ & $\kappa=1$ & $\kappa=0$ & $\kappa=-1 $ \\
\hline \hline 
+ & KS|S-dS|KS & KS & KS \\ \hline
0 & KS|Sch & KS & KS \\ \hline
- & KS|AdS & KS|AdS & KS|AdS \\ \hline

\end{tabular}\\ 
\caption{Summary of the classical solutions. The ones to the left are the 
interior solutions, to the right are exterior solutions. We also 
use the term KS in all cases where the spatial topology can be written as 
$S^1\times T_g$. Other abbreviations: S-dS: The Schwarzschild-deSitter 
family, Sch: The Schwarzschild solution, AdS: Anti-deSitter (or quotients 
of).}
\end{table}

\begin{figure} 
\centering
\epsfig{figure=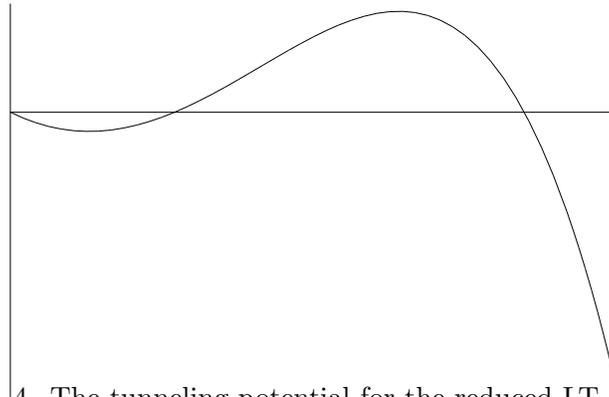, width=8.5cm} 
\caption{The tunneling potential for the reduced LT model}
\end{figure}

\begin{figure} 
\centering
\epsfig{figure=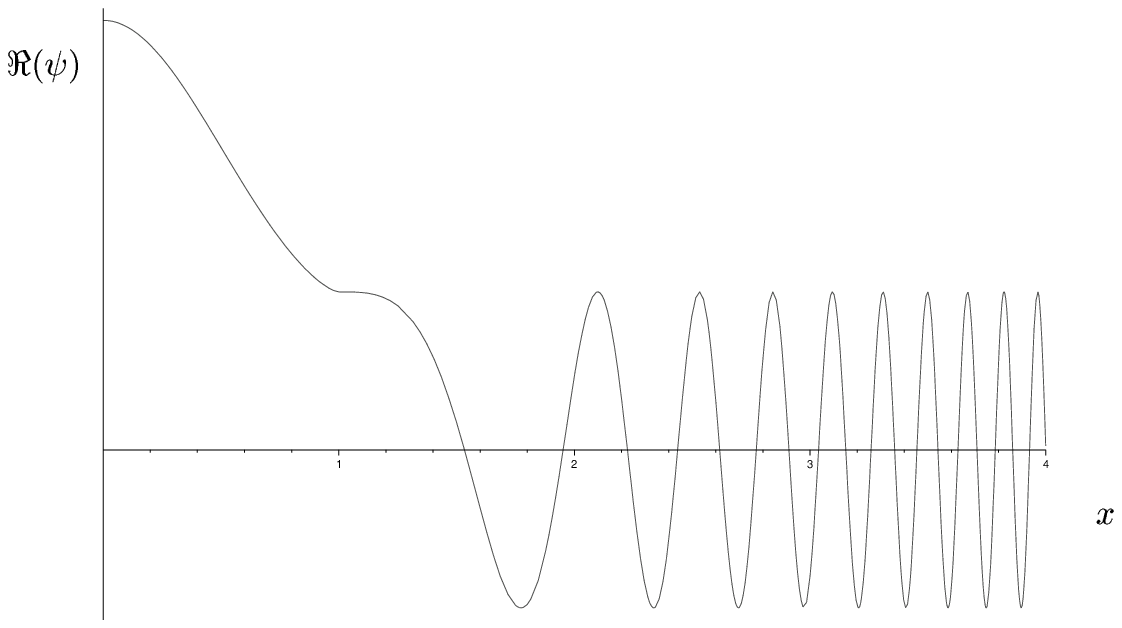, width=8.5cm}
\caption{The real part of the tunneling wavefunction of Vilenkin eq. 
\ref{viltun}}
\end{figure}


\begin{thebibliography}{99}
\bibitem{vil82}
A. Vilenkin, {\it Phys. Lett.} {\bf B117}, 25 (1982)
\bibitem{vil83}
A. Vilenkin, {\it Phys. Rev.} {\bf D27}, 2848 (1983)
\bibitem{vil84}
A. Vilenkin, {\it Phys. Rev.} {\bf D30}, 509 (1984)
\bibitem{vil86}
A. Vilenkin, {\it Phys. Rev.} {\bf D33}, 3560 (1986)
\bibitem{vv}
T. Vachaspati and A. Vilenkin, {\it Phys. Rev.} {\bf D37}, 898 (1988)
\bibitem{AP82}
D. Atkatz and H. Pagels, {\it Phys. Rev.}, {\bf D25}, 2065 (1982)
\bibitem{ll}
M. Lachi\`eze-Rey and J-P. Luminet, {\it Phys. Rep.} {\bf 254}, 135 (1995)
\bibitem{thur97}
W.P. Thurston, {\it Three-Dimensional Geometry and Topology}, 
{\bf Volume 1},  Princeton University Press (1997)
\bibitem{Kuchar}
K. Kucha\ck{r}, {\it Phys. Rev.} {\bf D50}, 3961 (1994)
\bibitem{KKTYa}
M. Kenmoku, H. Kubotani, E. Takasugi, Y. Yamazaki, {\it Phys. Rev.} 
{\bf D59}, (1999)
\bibitem{thurston}
W.P. Thurston, {\it Bull. Am. Math. Soc.} {\bf 6}, 357 (1982)
\bibitem{fugandes82}
H. Fagundes, {\it Lett. Math. Phys.} {\bf 6}, 417 (1982)
\bibitem{NS}
Y. Nambu and  M. Sasaki, {\it Prog. Theo. Phys.} {\bf 79}, 96 (1988)
\bibitem{KT}
K. Kucha\v{r} and C.G. Torre, {\it Phys. Rev. } {\bf D43}, 419 (1991)
\bibitem{salopek}
D.S. Salopek,  {\it Phys. Rev. } {\bf D46}, 4373 (1992)
\bibitem{hovedoppg}
S. Hervik, {\it Cand. Scient. Thesis}, University of Oslo (1999)
\bibitem{blp}
D. Brill, J. Louko and P. Peld\'an, {\it Phys. Rev.} {\bf D56}, 3600 (1997)
\bibitem{BTZ}
M. Ba\~nados, C. Teitelboim and J. Zanelli, {\it Phys. Rev. Lett.} 
{\bf 69} (1992)
\bibitem{SS}
D.D. Sokoloff and A.A. Starobinskii, {Sov. Astron.} {\bf 19}, 629 (1975)
\bibitem{sig}
S. Hervik, {\it Class. Quantum Grav.} {\bf 17}, 2765 (2000)
\bibitem{BH}
I. Bengtsson and S. Holst, {\it Class. Quantum Grav.} {\bf 16}, 3735 (1999)
\bibitem{Kra:97}
A. Krasi\'nski, {\it Inhomogeneous Cosmological models}, Cambridge 
University Press (1997)
\bibitem{Zecca:90}
A. Zecca, {\it Il nuovo Cimento}, {\bf 106B}, 413 (1990)
\bibitem{EG:95}
E. Eriksen and \O. Gr\o n, {\it Int. J. Mod. Phys.} {\bf 4}, 115 (1995)
\bibitem{KKTYb}
M. Kenmoku, H. Kubotani, E. Takasugi, Y. Yamazaki, gr-qc/9906056 (1999)
\bibitem{FMP:90}
W. Fischler, D. Morgan and J. Polchinski, {\it Phys. Rev.} {\bf D42}, 
4042 (1990)
\bibitem{Vilenkin88}
A. Vilenkin, {\it Phys. Rev.} {\bf D37}, 888 (1988)
\bibitem{Norbury97}
J.W. Norbury, {\it Eur. J. Phys.} {\bf B433}, 263 (1998)
\bibitem{Norbury98}
J.W. Norbury, {\it Phys. Lett.} {\bf 19}, 143 (1997)
\bibitem{CM99}
D.H. Coule and J. Martin, {\it Phys. Rev.} {\bf D61}, 063501 (2000)
\bibitem{con1}
H-D. Conradi, {\it Class. Quantum Grav.} {\bf 12}, 2423 (1995)
\bibitem{con2}
H-D. Conradi, {\it Int. J. Mod. Phys.} {\bf D7}, 189 (1998)
\bibitem{as}
A. Ashtekar and J. Samuel, {\it Class. Quantum Grav.} {\bf 8}, 2191 (1991)
\bibitem{kodama}
H. Kodama,  {\it Prog. Theo. Phys.} {\bf 99}, 173 (1998)
\end{thebibliography}
\end{document}